# Measurement of the Spin-Transfer-Torque Vector in Magnetic Tunnel Junctions

J. C. Sankey, Y.-T. Cui, R. A. Buhrman, D. C. Ralph*

*Cornell University, Ithaca, NY, 14853 USA*

J. Z. Sun, J. C. Slonczewski[†]

*IBM T. J. Watson Research Center, Yorktown Heights, NY 10598 USA*

Spin-polarized currents can transfer spin angular momentum to a ferromagnet, generating a torque that can efficiently reorient its magnetization. Achieving quantitative measurements of the spin-transfer-torque vector in magnetic tunnel junctions (MTJs) is important for understanding fundamental mechanisms affecting spin-dependent tunneling, and for developing magnetic memories and nanoscale microwave oscillators. Here we present direct measurements of both the magnitude and direction of the spin torque in $Co_{60}Fe_{20}B_{20}/MgO/Co_{60}Fe_{20}B_{20}$ MTJs. At low bias $V$, the differential torque $d\vec{\tau}/dV$ lies in the plane defined by the electrode magnetizations, and its magnitude is in excellent agreement with a prediction for highly-spin-polarized tunneling. With increasing bias, the in-plane component $d\tau_{\parallel}/dV$ remains large, in striking contrast to the decreasing magnetoresistance ratio. The differential torque vector also rotates out of the plane under bias; we measure a perpendicular component $\tau_{\perp}(V)$ with bias dependence $\propto V^2$ for low $V$, that becomes as large as 30% of the in-plane torque.

*email: ralph@ccmr.cornell.edu

[†]IBM RSM Emeritus

Magnetic tunnel junctions (MTJs) with MgO barriers can have extremely large magnetoresistance, and for this reason they are being pursued aggressively for applications in memory technologies and magnetic-field sensing[1,2,3,4]. Further, it has recently been demonstrated that the magnetic state of a nanoscale MTJ can be switched by a spin-polarized tunnel current via the so-called spin-transfer torque[5,6]. This is a promising new mechanism for the write operation of nanomagnetic memory elements[7] and for driving nanoscale microwave oscillators[8,9,10]. While the presence of the spin torque has been unambiguously observed, its quantitative behavior in an MTJ, especially its bias dependence, has yet to be understood in detail. One puzzling observation has been that, in contrast to the tunnel magnetoresistance, the spin torque has been found to depend very little on the junction bias[11]. Recent theoretical models attempt to quantify the spin torque's bias dependence in an MTJ, and to explain its relationship with the tunnel magnetoresistance[12,13,14,15]. To test these model calculations, a direct, quantitative measurement of how the spin-torque varies with junction bias is highly desirable. Quantitative understanding of this bias dependence will also be important for the development and optimization of nanostructured MTJ spin-torque devices in memory applications.

Here we use the recently-developed technique of spin-transfer-driven ferromagnetic resonance (ST-FMR)[16,17] to measure the bias- and angular-dependence of the spin-transfer torque $\vec{\tau}$ in MgO-based junctions. We demonstrate for the first time that ST-FMR can be used to achieve a detailed, highly-quantitative understanding of spin torque in individual nanoscale devices. We define $\hat{m}$ and $\hat{M}_{fixed}$ as unit vectors in the direction of the magnetic moments on the two sides of our magnetic tunnel junction and $V$ as the bias voltage. Our measurements show that at low bias the in-plane component (in the direction $\hat{m} \times (\hat{m} \times \hat{M}_{fixed})$) of the spin transfer "torkance" $d\tau_{\parallel}/dV$ on moment $\hat{m}$ is equal, within the experimental accuracy (±15%), to the value predicted for an elastic tunnel current with a spin polarization $P = 0.66$ appropriate for our junctions[18]. For $P = 1$, the predicted in-plane torque is only 8% stronger than for $P = 0.66$, from which we conclude that producing MTJ's for spin-torque applications with higher $P$ will not yield a substantial further improvement in the strength of the torque. We find the in-plane torkance to be essentially bias-independent, varying less than 8% below 0.3 V, and appearing to increase slightly at even higher bias where sample heating may start to affect measurements. In contrast, the magnetoresistance decreases by 72% over our bias range. We also measure a perpendicular component $d\tau_{\perp}/dV$ (in the $\hat{m} \times \hat{M}_{fixed}$ direction) that is nonzero only when $V \neq 0$, with a bias-dependence corresponding to $\tau_{\perp}(V) = A_0 + A_1 V^2$ near $V=0$ (with $A_0$ and $A_1$ constants). This perpendicular component could have a substantial effect in improving spin-torque-driven magnetic reversal in MTJs. Our measurements can be interpreted within a simple model.

We have studied 8 exchange-biased tunnel junctions with the layers (in nm) 5 Ta / 20 Cu / 3 Ta / 20 Cu / 15 PtMn / 2.5 $Co_{70}Fe_{30}$ / 0.85 Ru / 3 $Co_{60}Fe_{20}B_{20}$ / 1.25 MgO / 2.5 $Co_{60}Fe_{20}B_{20}$ / 5 Ta / 7 Ru deposited on an oxidized silicon wafer by the process described in ref. [19] (See Fig. 1a). The top magnetic layer (with moment direction $\hat{m}$) is etched to a rounded rectangular cross section with the long axis parallel to the exchange bias from the PtMn layer (the $\hat{y}$ direction), with size either 50 × 100 nm² or 50 × 150 nm². The bottom layer (moment direction $\hat{M}_{fixed}$) is left extended on the scale of 10's of microns. All data in this paper are from one 50 × 100 nm² device; the other samples gave similar behavior. Contact pads are originally fabricated in a 4-point configuration, but we cut the top electrode close to the sample (Fig. 1b, left inset) prior to ST-FMR measurements to eliminate artifacts associated with RF current flow within this electrode (between contact pads A and B) rather than through the tunnel junction. The bias dependence of the differential resistance $dV/dI$ is shown in Fig. 1b for the parallel magnetization orientation (P, $\theta = 0°$, with $\theta$ the angle between $\hat{m}$ and $\hat{M}_{fixed}$), antiparallel (AP, $\theta = 180°$), and intermediate angles. At zero bias, the tunneling magnetoresistance ratio (TMR) is $[(dV/dI)_{AP} - (dV/dI)_P]/[(dV/dI)_P] = 154\%$. The TMR decreases to 43% at 540 mV bias, a fractional reduction of 72%.

The ST-FMR measurements[16,17] are performed at room temperature. We apply a sufficiently-strong magnetic field $H$ along the $\hat{z}$ direction (Fig. 1b inset) to saturate $\hat{m}$, while $\hat{M}_{fixed}$ is tilted to a lesser degree away from $\hat{y}$. Representative results for the ST-FMR spectra are shown in Fig. 2. We observe several magnetic resonances in the frequency range 2 to 14 GHz. The lowest-frequency resonance has the largest amplitude, and corresponds to the sign of the signal expected for excitation of the free magnetic layer[17]. We assume that other smaller resonances correspond to higher-frequency standing-wave modes of the free or fixed layer, or perhaps coupled modes[17].

Our first major result is that the degree of asymmetry in the ST-FMR peak shape *vs.* frequency for the lowest-frequency mode depends strongly on dc bias current *I*, with peak shapes for *I*=0 being symmetric, and with the sign of the asymmetry depending on the sign of *I* (Fig. 2b). To analyze quantitatively the magnitudes and the peak shapes of the ST-FMR signals, we assume that the dynamics of the free magnetic layer near the main resonance peak can be described by a simple macrospin approximation, so that a generalized Landau-Lifshitz-Gilbert (LLG) equation applies:

$$\frac{d\hat{\mathbf{m}}}{dt} = -\gamma \hat{\mathbf{m}} \times \vec{\mathbf{H}}_{\text{eff}} + \alpha \hat{\mathbf{m}} \times \frac{d\hat{\mathbf{m}}}{dt} - \gamma \frac{\tau_{\parallel}(I,\theta)}{M_s Vol} \hat{\mathbf{y}} - \gamma \frac{\tau_{\perp}(I,\theta)}{M_s Vol} \hat{\mathbf{x}}. \qquad (1)$$

Here $\gamma$ is the magnitude of the gyromagnetic ratio, $\alpha$ is the Gilbert damping parameter, $\vec{\mathbf{H}}_{\text{eff}}$ is an effective field as defined in ref. [20], and $M_s Vol \approx (1.06 \pm 0.16) \times 10^{-14}$ emu is the total magnetic moment of the free layer based on our estimate of the sample geometry and the measured value of $M_s = 1100$ emu/cm$^3$, consistent with ref. [19]. The resulting ST-FMR lineshapes have been evaluated[16,20,21], and good agreement has been observed in ST-FMR measurements on all-metal spin-valve devices[17]. By extending the analysis of ref. [20] to nonzero values of $I$ (see Supplementary Note 1), this formalism predicts that the ST-FMR signal is to a good approximation

$$\langle V_{\text{mix}} \rangle = \frac{1}{4} \frac{\partial^2 V}{\partial I^2} I_{RF}^2 + \frac{1}{2} \frac{\partial^2 V}{\partial \theta \partial I} \frac{\hbar \gamma \sin \theta}{4e M_s Vol \, \sigma} I_{RF}^2 \left( \zeta_{\parallel} S(\omega) - \zeta_{\perp} \Omega_{\perp} A(\omega) \right). \qquad (2)$$

Here $\zeta_{\parallel} = [(2e/\hbar)/\sin(\theta)] \, d\tau_{\parallel}/dI$ and $\zeta_{\perp} = [(2e/\hbar)/\sin(\theta)] \, d\tau_{\perp}/dI$ represent the differential torques in dimensionless units, $S(\omega) = 1/\{1 + [(\omega - \omega_m)/\sigma]^2\}$ and $A(\omega) = [(\omega - \omega_m)/\sigma] S(\omega)$ are symmetric and antisymmetric Lorentzians, $\sigma$ is the linewidth, $\omega_m$ is the resonant precession frequency, and $\Omega_{\perp} = \gamma(4\pi M_{\text{eff}} + H)/\omega_m$ for our geometry. We use $4\pi M_{\text{eff}} = 11 \pm 1$ kOe for the effective out-of-plane anisotropy, as determined from the magnetoresistance for $H$ perpendicular to the substrate. The first term on the right in Eq. (2) is a non-resonant background, useful for calibrating $I_{RF}$. The second term gives the dominant ST-FMR signal; as a function of frequency it has the form of a symmetric Lorentzian $\propto \zeta_{\parallel} \propto d\tau_{\parallel}/dI$, minus an antisymmetric Lorentzian $\propto \zeta_{\perp} \propto d\tau_{\perp}/dI$.

As shown in Fig. 2b, the peak shapes for the ST-FMR signals of the lowest-frequency main resonance mode are fit very well by the form expected from Eq. (2). From the fits, at each value of $H$ and $I$ we determine with high precision the symmetric and antisymmetric peak amplitudes, the background, the linewidth $\sigma$, and the resonant frequency $\omega_m$. The raw results of these fits are shown in Supplementary Figure S1. In order to make a quantitative determination of $d\tau_\parallel/dI$ and $d\tau_\perp/dI$ using Eq. (2), it is necessary to calibrate the quantities $I_{RF}^2$ and $\partial^2 V/\partial\theta\partial I$, both of which depend on $I$ due to the bias dependence of the tunnel-junction impedance. We determine $I_{RF}^2$ from the non-resonant background signal, together with the value of $\partial^2 V/\partial I^2$ determined at low frequency. We calibrate $\partial^2 V/\partial\theta\partial I$ by measuring $\partial V/\partial I$ vs. $I$ at a sequence of magnetic fields in the $\hat{z}$ direction, assuming that the zero-bias conductance varies as $\cos(\theta)$ (and that $\theta$ depends negligibly on $I$), and then numerically differentiating $\partial V/\partial I$ with respect to $\theta$ at each value of $I$ and $H$. These calibrations are sufficiently accurate that the uncertainty in our measurements is dominated by the uncertainty in the determination of $M_s Vol$, not $I_{RF}$ or $\partial^2 V/\partial\theta\partial I$ or the quality of fits to the peak shapes. Additional details concerning the calibration procedures are given in Supplementary Note 2 and Supplementary Figures S2 and S3.

The most relevant final quantities for physical interpretation are expected to be the "torkances"[18], $d\tau_\parallel/dV = (d\tau_\parallel/dI)/(dV/dI)$ and $d\tau_\perp/dV = (d\tau_\perp/dI)/(dV/dI)$. We plot these in Fig. 3a, as calculated from the measured values of $dV/dI$ and the values of $d\tau_\parallel/dI$ and $d\tau_\perp/dI$ determined from the second term on the right side of Eq. (2). ($d\tau_\parallel/dI$ and $d\tau_\perp/dI$ are plotted in Supplementary Figure S4 and the ratio $(d\tau_\perp/dV)/(d\tau_\parallel/dV)$ in Supplementary Figure S5.) We first consider the dependence of the torkances on $\theta$. It is predicted[12,13,18] that for elastic tunneling $d\vec{\tau}/dV$ should be $\propto \sin(\theta)$. The inset to Fig. 3a shows that $(d\tau_\parallel/dV)/\sin(\theta)$ is indeed nearly constant over the range of angles measured, $45° < \theta < 90°$. Given this agreement, we divide out a factor of $\sin(\theta)$ in plotting the torkances in the main panel of Fig. 3a, so that the plotted results should be independent of angle. ($\theta$ is determined as discussed above.)

The bias dependence of the dominant, in-plane component of the torkance, $d\tau_\parallel/dV$, is shown in the main panel of Fig. 3a. At $V=0$, we find $(d\tau_\parallel/dV)/\sin(\theta) = 0.13 \pm 0.02$ $\hbar/(2e)$ k$\Omega^{-1}$. The value predicted[18] for elastic tunneling in a symmetric junction of polarization $P$ is $(d\tau_\parallel/dV)/\sin\theta = [\hbar/(4e)][2P/(1+P^2)](dI/dV)_P$, which is equal to 0.144 $\hbar/(2e)$ k$\Omega^{-1}$ for $P = 0.66$ (corresponding to our TMR of 154%) and the value of parallel conductance $(dI/dV)_P = 1/(3.19$ k$\Omega)$ for our device. For $P=1$, the prediction would be 0.157 $\hbar/(2e)$ k$\Omega^{-1}$. Therefore our measured torkance agrees with this prediction to within experimental uncertainty, and is within 20% of the maximum value possible given our device conductance. Consequently, attempts to produce MTJ's for spin torque applications that have even higher TMR values are unlikely to improve the spin-torque-to-conductance ratio by more than this small amount. The reason why the TMR and torkance are not more closely linked is that inelastic tunneling due to magnons[15], and generally other mechanisms not involving spin operators[18], may decrease TMR without affecting torkance in symmetric MTJs.

As a function of bias, we find that $d\tau_\parallel/dV$ is constant to within ±8% for $|V| \leq 300$ mV. This is in striking contrast to the magnetoresistance, which decreases by 50% over the same bias range (Fig. 1a). Furthermore, the value of $d\tau_\parallel/dV$ appears to increase for 300 mV < |V| < 540 mV, whereas the magnetoresistance continues to decrease to just 28% of its full value. The low-bias result confirms with greater sensitivity the conclusions in ref. [11], in which a combined effect of $\tau_\parallel/I$ and $\tau_\perp/I$ was measured for |V| < 350 mV in $Co_{90}Fe_{10}/MgO/Co_{90}Fe_{10}$ junctions.

The theoretical framework of ref. [18] provides a means to analyze these results. The differential conductances for parallel and antiparallel magnetic configurations and the in-plane spin-transfer torkance can all be written in terms of conductance amplitudes $G_{\sigma\sigma'}$ between spin channels ($\sigma, \sigma' = \pm$ are spin indices for the bottom and top electrodes). Assuming that the tunneling mechanism itself does not depend on spin operators, we may write[13,18]

$$\frac{d\tau_\parallel}{dV} = \frac{\hbar}{4e}(G_{++} - G_{--} + G_{+-} - G_{-+})\sin(\theta) \tag{3}$$

$$\left(\frac{dI}{dV}\right)_P = G_{++} + G_{--}, \quad \left(\frac{dI}{dV}\right)_{AP} = G_{+-} + G_{-+} \tag{4}$$

The amplitudes $G_{\sigma\sigma'}$ can describe both elastic and inelastic tunneling processes. With the assumptions that $G_{+-} \approx G_{-+}$ for a symmetric junction near zero bias and $G_{--} \ll G_{++}$, these equations imply that, approximately, $d\tau_\parallel/dV \propto (dI/dV)_P$. The observation that $d\tau_\parallel/dV$ is approximately independent of bias for |V| < 300 mV can therefore be related to the fact that the differential conductance for parallel moments is approximately independent of bias in this range, as well. Figure 3b shows a direct comparison of the fractional changes in $d\tau_\parallel/dV$ and $(dI/dV)_P$ vs. V. For |V| < 300 mV, $d\tau_\parallel/dV$ and $(dI/dV)_P$ display a similar pattern of non-monotonic variations, although the relative

changes in $d\tau_\parallel/dV$ are greater. At larger biases, 300 mV < |V| < 540 mV, the apparent experimental value of $d\tau_\parallel/dV$ increases much more rapidly than $(dI/dV)_P$. One possible explanation for this upturn may be heating. Previous studies of magnetic tunnel junctions[11,22] suggest that the effective temperature of our free layer may be heated 50-100 K or more above room temperature at our highest biases. This could decrease the total magnetic moment of the free layer ($M_s Vol$) thereby enhancing the response of the magnet to a given torkance and artificially inflating our determination of $d\vec{\tau}/dV$ for |V| > 300 mV.

Within the macrospin ST-FMR model (Eq. (2)), an antisymmetric-in-frequency component of the ST-FMR resonance can be related to an out-of-plane torkance, $d\tau_\perp/dV$. We observe only symmetric ST-FMR peaks at V=0 (Fig. 2b), implying that at zero bias $d\tau_\perp/dV=0$. This differs from a previous experimental report[16]. Fig. 3a shows that the asymmetries we measure for $V \neq 0$ correspond to an approximately linear dependence of $d\tau_\perp/dV$ on V at low bias. This result is consistent with theoretical expectations[13,14] that the low-order bias dependence has the form $\tau_\perp(V)/\sin(\theta) = A_0 + A_1 V^2$ (with $A_0$ and $A_1$ independent of bias). For our full range of bias we measure $A_1 = (84 \pm 13)(\hbar/2e)\,G\Omega^{-1}V^{-1}$. The integrated torque $\tau_\perp(V)$ is in the $\hat{m} \times \hat{M}_{fixed}$ direction, and grows to be 30% of the in-plane torque $\tau_\parallel(V)$ at the largest bias we probe. This perpendicular torque may be important for magnetic memory applications, because a component of torque in this direction may reduce the current needed for switching[23]. We do not believe that alternative mechanisms such as heating can account for the asymmetric peak shapes, as explained in Supplementary Note 3.

We have also performed ST-FMR measurements on metallic IrMn / Py / Cu / Py spin valves in the same experimental geometry, and in that case we find that the lowest-frequency peaks are frequency-symmetric to within experimental accuracy for all biases $|I| < 2$ mA, from which we conclude that $\tau_\perp$ is always less than 1% of $\tau_\parallel$ (see Supplementary Figure S6). The ratio $\tau_\perp / \tau_\parallel < 1\%$ is much smaller than has been suggested for Co/Cu/Co metal spin valves based on analysis of the dynamical phase diagram[24]. We suspect that the existence of a significant perpendicular component of the spin torque in current-perpendicular-to-the-plan multilayer devices is particular to tunnel junctions.

The measured linewidths $\sigma$ of our ST-FMR measurements on MgO junctions allow a determination of the magnetic damping. Within the macrospin model (Supplementary Note 1), assuming that $\tau_\parallel(V,\theta) \propto \sin(\theta)$,

$$\sigma = \frac{\alpha \omega_m}{2}\left(\Omega_\perp + \Omega_\perp^{-1}\right) - \cot(\theta)\frac{\gamma \tau_\parallel(V,\theta)}{2M_s Vol}. \tag{5}$$

In Fig. 3c we plot the bias dependence of the effective damping defined as $\alpha_{eff} = 2\sigma/[\omega_m(\Omega_\perp + \Omega_\perp^{-1})]$. The zero-bias values give an average Gilbert damping coefficient $\alpha = 0.0095 \pm 0.0010$, consistent with literature reports for similar materials[25]. The lines plotted in Fig. 3c show the slopes expected from Eq. (5), using as a fitting parameter that $\partial \tau_\parallel / \partial V / (\sin(\theta)) = (0.16 \pm 0.03)$ k$\Omega^{-1} \hbar/2e$ (assuming that $\partial \tau_\parallel / \partial V$ is constant for $|V| < 300$ mV). This estimate agrees with the value determined independently above from the magnitude of the ST-FMR peak.

In summary, we have employed spin-transfer-driven FMR to achieve detailed quantitative measurements of the spin-transfer torque and magnetic damping in individual $Co_{60}Fe_{20}B_{20}/MgO/Co_{60}Fe_{20}B_{20}$ magnetic tunnel junctions, of the type that are of interest for nonvolatile magnetic random access memory applications. We find that the dominant, in-plane component $d\tau_\parallel / dV$ has a magnitude at zero bias equal to, within the experimental uncertainty of 15%, the value predicted for elastic tunneling in a symmetric junction. The torkance $d\tau_\parallel / dV$ is independent of bias to within ±8% for |V| ≤ 300 mV, and shows no evidence of weakening even at higher bias. The observations that $d\tau_\parallel / dV$ in MgO tunnel junctions is in good agreement with the predicted magnitude, close to the maximum possible value expected for fully spin-polarized elastic tunneling, and that this torkance maintains its strength at high bias where the magnetoresistance is small, have important consequences for the development of spin-transfer-switched memory devices. We also observe for the first time a bias-dependent perpendicular component of the torque in magnetic tunnel junctions with, to a good approximation, $\tau_\perp(V) \propto A_0 + A_1 V^2$ at low bias, which agrees with predictions. This component of the torque is sufficiently strong at high bias that it could assist efficient magnetic reversal, and it should be included in device modeling.

## METHODS

### DEVICE FABRICATION

The resistance-area product for parallel electrodes in our tunnel-junction multilayers is ≈12 $\Omega\mu m^2$. The top $Co_{60}Fe_{20}B_{20}$ layer is patterned by electron beam lithography and ion milling to produce a rectangular cross section with rounded corners. The etch is stopped at the MgO barrier. Top contacts are made with 5 nm Ti / 150 nm Cu / 10 nm Pt. The contacts are originally fabricated in a 4-probe arrangement as shown in Fig. 1b.

However, we find that ST-FMR measurements in this geometry are affected by RF current flowing within the patterned top electrode above the sample (rather than through the MgO tunnel barrier). This produces a significant RF magnetic field with a phase different than the spin current, and changes the magnitude and the degree of asymmetry of the resonance peak. It also causes the FMR results to vary depending on which of the two top contacts (A or B in Fig. 1b) is used, while the results are the same upon interchanging bottom contacts (C or D). Similar effects from RF currents flowing past the tunnel junction may also have affected a previous ST-FMR measurement of MgO devices[16], which showed significantly asymmetric line shapes even at zero dc bias. To minimize this problem, we cut the top lead near the sample (see Fig. 1b), and then perform the ST-FMR measurements using contacts B and D.

ST-FMR MEASUREMENTS

Our ST-FMR measurements are conducted using the procedure described in ref. [17]. A direct current $I$ and a microwave-frequency current $I_{RF}$ are applied simultaneously to the sample at room temperature via a bias-tee. When spin-transfer from $I_{RF}$ excites resonant magnetic dynamics, the resulting resistance oscillations mix with $I_{RF}$ to produce a DC voltage response $V_{mix}$. To maximize the signal-to-noise of the measurement, we chop $I_{RF}$ at 250 Hz and measure $V_{mix}$ using a lock-in amplifier. In all cases, we use values of $I_{RF}$ in the range 5-25 µA, small enough that the FMR response is in the linear regime. Our procedures for calibrating $I_{RF}$ and preventing variations in $I_{RF}$ while sweeping frequency are discussed in Supplementary Note 2. We use the convention that positive bias corresponds to electron flow down the pillar, giving a sign of $\tau_\parallel$ that favors antiparallel alignment of the top "free" magnetic layer moment ($\hat{m}$) relative to that of the lower layer

($\hat{\boldsymbol{M}}_{fixed}$).

Acknowledgements

We thank Y. Nagamine, D. D. Djayaprawira, N. Watanabe, and K. Tsunekawa of Canon ANEVA Corp. for providing the junction thin film stack which we used to fabricate the tunnel junction devices in this study, and G. D. Fuchs and K. V. Thadani for discussions. JZS would like to acknowledge fruitful discussions with S. Assefa, W. J. Gallagher on sample processing techniques, X. Jiang and S. S. P. Parkin for invaluable sample fabrication assistance, and the support of the IBM MRAM team as a whole. Cornell acknowledges support from the Office of Naval Research, from the NSF (DMR-0605742), and from the NSF/NSEC program through the Cornell Center for Nanoscale Systems. We also acknowledge NSF support through use of the Cornell Nanofabrication Facility/NNIN and the Cornell Center for Materials Research facilities.

Competing Interests Statement

The authors declare that they have no competing financial interests.

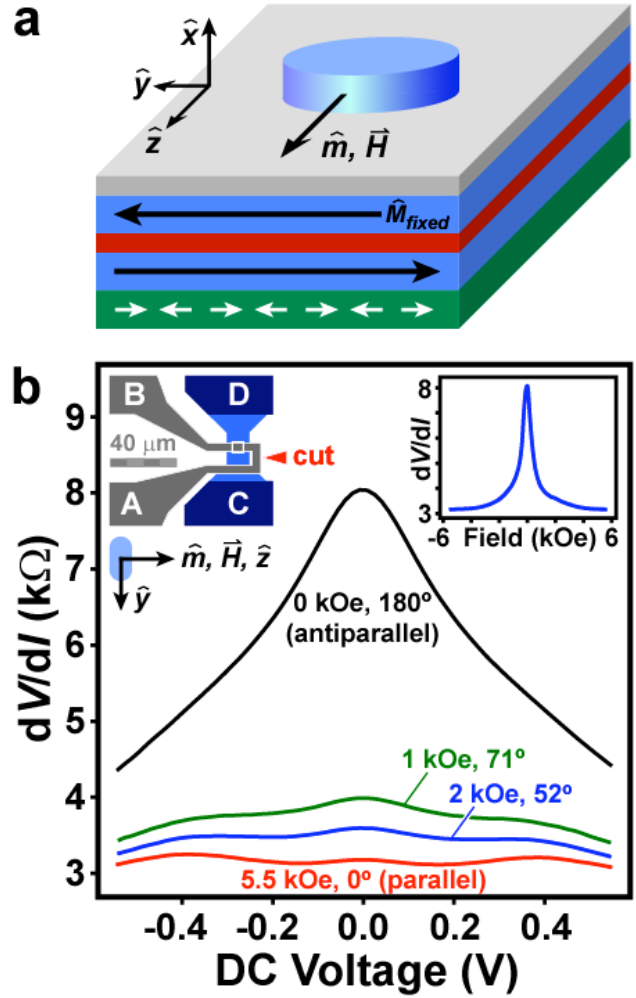

Fig. 1. **Magnetic tunnel junction geometry and magnetic characterization. a** Schematic of the sample geometry. **b** Bias dependence of differential resistance at room temperature for the parallel orientation of the magnetic electrodes ($\theta = 0°$) and antiparallel orientation ($\theta = 180°$), along with intermediate angles. The angles are determined assuming that the zero-bias conductance varies as $\cos(\theta)$. (Left inset) Layout of the electrical contacts (cropped), showing where the top electrode is cut to eliminate measurement artifacts. (Right inset) Zero-bias magnetoresistance for $H$ along $\hat{z}$.

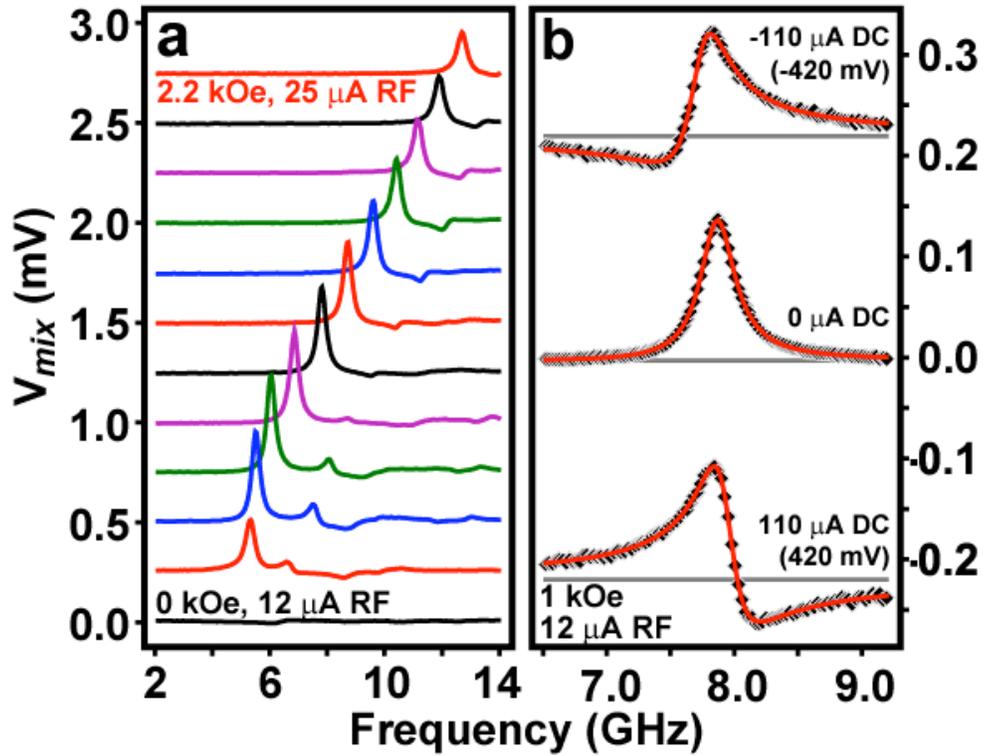

Fig. 2. **ST-FMR spectra at room temperature.** **a** Spin-transfer FMR spectra for $I = 0$, for magnetic fields (along $\hat{z}$) spaced by 0.2 kOe. $I_{RF}$ ranges from 12 µA at low field (high resistance) to 25 µA at high field. The curves are offset by 250 µV. **b** Details of the primary ST-FMR peaks at $H = 1000$ Oe and $I_{RF} \approx 12$ µA for different DC biases. Symbols are data, lines are Lorentzian fits. These curves are not artificially offset; the frequency-independent backgrounds for nonzero DC biases correspond to the first term on the right of Eq. (2). A DC bias also changes the degree of asymmetry in the peak shape vs. frequency.

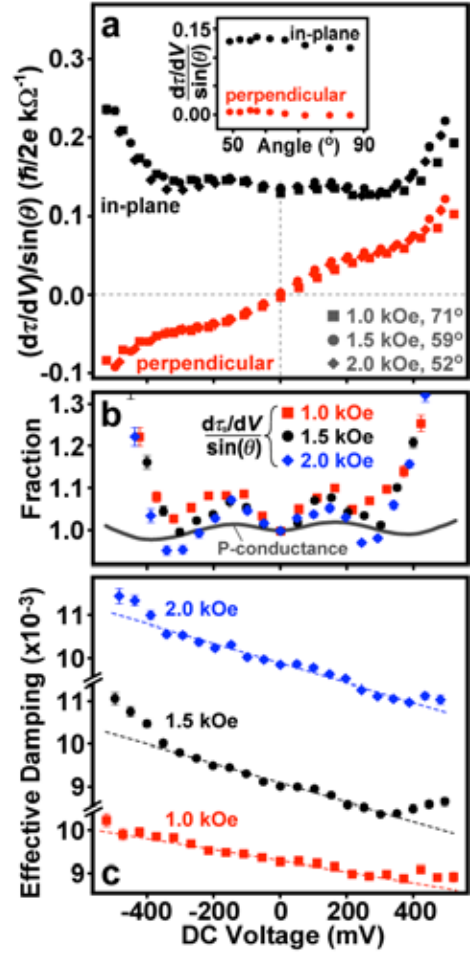

Fig. 3. **Bias dependence of the spin-transfer torkances and magnetic damping. a** Magnitudes of the in-plane torkance $d\tau_\parallel/dV$ and the out-of-plane torkance $d\tau_\perp/dV$ determined from the room temperature ST-FMR signals, for three different values of applied magnetic field in the $\hat{z}$ direction. The overall scale for the torkances has an uncertainty of ~15% associated with the determination of the sample volume. (Inset) Angular dependence of the torkances at zero bias. **b** Comparison of the bias dependences of $d\tau_\parallel/dV$ and $(dI/dV)_P$, relative to their zero-bias values. Small background slopes (visible in **a**) are subtracted from the torkance values. **c** Symbols: Effective damping determined from the ST-FMR linewidths. Lines: Fit to Eq. (5).

# Supplementary Material

Supplementary Note 1: Derivation of the ST-FMR signal $\langle V_{mix} \rangle$ (Eq. (2) in the main text)

This derivation generalizes arguments in references [S1,S2,S3] in order to consider experiments in which a finite bias is applied to the sample.

We consider only the specific geometry relevant to our experiment and define the coordinate axes as in ref. [S2]. We assume that the orientation $\hat{m}$ of the free-layer moment undergoes small-angle precession about the $\hat{z}$ axis, that the plane of the sample is the $\hat{y}$-$\hat{z}$ plane, that easy axis of the free layer is along $\hat{y}$, and that the orientation $\hat{M}_{fixed}$ of the fixed-layer moment is in the plane of the sample and differs from $\hat{z}$ by an angle $\theta_0$ toward $\hat{y}$. Let $\theta$ be the angle between $\hat{m}$ and $\hat{M}_{fixed}$. The precession of the free layer in response to the current $I(t) = I + \delta I(t)$ (where $\delta I(t) = I_{RF} \text{Re}(e^{i\omega t})$) can be characterized by the transverse components $m_x(t) = \text{Re}(m_x e^{i\omega t})$ and $m_y(t) = \text{Re}(m_y e^{i\omega t})$. Because of the large magnetic anisotropy of the thin film sample, $|m_x| << |m_y|$ and changes in the angle $\theta$ during precession are to good approximation $\delta\theta(t) = -\text{Re}(m_y e^{i\omega t})$.

The time-dependent voltage $V(t)$ across the sample will depend on the instantaneous value of the current and $\theta$. The DC voltage signal produced by rectification in ST-FMR can be calculated by Taylor-expanding $V(t)$ to 2$^{nd}$ order and taking the time average over one precession period

$$\langle V_{mix} \rangle = \frac{1}{2} \frac{\partial^2 V}{\partial I^2} \langle (\delta I(t))^2 \rangle + \frac{\partial^2 V}{\partial I \partial \theta} \langle (\delta I(t))(\delta\theta(t)) \rangle + \frac{1}{2} \frac{\partial^2 V}{\partial \theta^2} \langle (\delta\theta(t))^2 \rangle. \tag{S1}$$

Here $\langle \rangle$ denotes the time average. With this expression, we assume that voltage signals due to spin pumping [S3] are negligible in tunnel junctions. Using $\delta\theta(t) = -\text{Re}(m_y e^{i\omega t})$, Eq. (S1) can be expressed

$$\langle V_{mix} \rangle = \frac{1}{4} \frac{\partial^2 V}{\partial I^2} I_{RF}^2 - \frac{1}{2} \frac{\partial^2 V}{\partial I \partial \theta} I_{RF} \text{Re}(m_y) + \frac{1}{4} \frac{\partial^2 V}{\partial \theta^2} |m_y|^2. \tag{S2}$$

We calculate the precession angle $m_y$ from the Landau-Lifshitz-Gilbert equation of motion in the macrospin approximation, with the addition of spin-transfer-torque terms transverse to the free-layer moment.

$$\frac{d\hat{\mathbf{m}}}{dt} = -\gamma \hat{\mathbf{m}} \times \vec{H}_{eff} + \alpha \hat{\mathbf{m}} \times \frac{d\hat{\mathbf{m}}}{dt} - \gamma \frac{\tau_\parallel(I,\theta)}{M_s Vol} \hat{y} - \gamma \frac{\tau_\perp(I,\theta)}{M_s Vol} \hat{x}, \tag{S3}$$

where $\gamma$ is the magnitude of the gyromagnetic ratio, $\alpha$ is the Gilbert damping coefficient, and $M_s Vol$ is the total magnetic moment of the free layer. For our specific experimental geometry, $\vec{H}_{eff} = -N_x M_{eff} \hat{x} - N_y M_{eff} \hat{y}$ with $N_x = 4\pi + (H/M_{eff})$ and $N_y = (H - H_{anis})/M_{eff}$. Here $H$ is the external magnetic field along $\hat{z}$, $4\pi M_{eff}$ is the effective anisotropy perpendicular to the sample plane, and $H_{anis}$ denotes the strength of anisotropy within the easy plane. (If the precession axis is not along a high-symmetry direction like $\hat{z}$, there are additional off-diagonal demagnetization terms in $\vec{H}_{eff}$ that will make the general expression for the ST-FMR signal more complicated than the one that we derive here [S2].)

For small RF excitation currents, the spin-torque terms can be Taylor-expanded,

$$\tau_\parallel(I,\theta) = \tau_\parallel^0 + \frac{\partial \tau_\parallel}{\partial I} \delta I(t) + \frac{\partial \tau_\parallel}{\partial \theta} \delta\theta(t), \quad \tau_\perp(I,\theta) = \tau_\perp^0 + \frac{\partial \tau_\perp}{\partial I} \delta I(t) + \frac{\partial \tau_\perp}{\partial \theta} \delta\theta(t). \tag{S4}$$

We have used a different sign convention than ref. [S2], so that the variables $\eta_1$ and $\eta_2$ in ref. [S2] correspond at zero bias to $\eta_1 = -\frac{2e}{\hbar \sin(\theta)} \frac{\partial \tau_\parallel}{\partial I} \equiv -\zeta_\parallel$ and $\eta_2 = -\frac{2e}{\hbar \sin(\theta)} \frac{\partial \tau_\perp}{\partial I} \equiv -\zeta_\perp$ in our notation.

The oscillatory terms in the equation of motion are

$$i\omega m_x = -m_y(\gamma N_y M_{eff} + i\alpha\omega) - \frac{\gamma}{M_s Vol}\left(\frac{\partial \tau_\perp}{\partial I} I_{RF} - \frac{\partial \tau_\perp}{\partial \theta} m_y\right).$$

$$i\omega m_y = m_x(\gamma N_x M_{eff} + i\alpha\omega) - \frac{\gamma}{M_s Vol}\left(\frac{\partial \tau_\parallel}{\partial I} I_{RF} - \frac{\partial \tau_\parallel}{\partial \theta} m_y\right).$$

(S5)

At this stage, we have neglected the influence of the DC spin-torque terms in shifting the precession axis of the free layer away from $\hat{z}$. For the bias range of our experiment, this is a very small effect. Solving these equations for $m_y$ to lowest order in the damping coefficient $\alpha$ we have

$$m_y = \frac{\gamma I_{RF}}{2 M_s Vol} \frac{1}{(\omega - \omega_m - i\sigma)}\left[i\frac{\partial \tau_\parallel}{\partial I} + \frac{\gamma N_x M_{eff}}{\omega_m}\frac{\partial \tau_\perp}{\partial I}\right].$$

(S6)

Here, the resonant precession frequency is $\omega_m = \gamma M_{eff}\sqrt{N_x N_y}$ and the linewidth is

$$\sigma = \frac{\alpha \gamma M_{eff}(N_x + N_y)}{2} - \frac{\gamma}{2 M_s Vol}\frac{\partial \tau_\parallel}{\partial \theta}.$$

(S7)

In the expression for the resonant precession frequency, we have neglected a correction $\propto \partial \tau_\perp / \partial \theta$ that is negligible for our experiment. The small shifts in the resonant frequency that we measure as a function of bias (see Supplementary Figure S3c) may be associated with micromagnetic phenomena that go beyond our macrospin approximation [4].

If we define $S(\omega) = 1/\{1 + [(\omega - \omega_m)/\sigma]^2\}$, $A(\omega) = [(\omega - \omega_m)/\sigma]S(\omega)$, and $\Omega_\perp = \gamma N_x M_{eff}/\omega_m$, and substitute Eq. (S6) into Eq. (S2), we reach:

$$\langle V_{mix}\rangle = \frac{1}{4}\frac{\partial^2 V}{\partial I^2}I_{RF}^2 + \frac{1}{2}\frac{\partial^2 V}{\partial\theta\partial I}\frac{\hbar\gamma\sin\theta}{4e\,M_s Vol\,\sigma}I_{RF}^2\left(\zeta_\parallel S(\omega) - \zeta_\perp \Omega_\perp A(\omega)\right)$$

$$+ \frac{1}{4}\frac{\partial^2 V}{\partial\theta^2}\left(\frac{\hbar\gamma\sin\theta}{4e\,M_s Vol\,\sigma}\right)^2 I_{RF}^2\left(\zeta_\parallel^2 + \zeta_\perp^2\Omega_\perp^2\right)S(\omega).$$

(S8)

The final term in Eq. (S8) represents a DC voltage generated by a change in the average low-frequency resistance due to magnetic precession. This term should be approximately an odd function of bias, and we estimate that it is small in the bias range we explore. It may be the explanation for the small slope in the dependence of $d\tau_\parallel/dV$ vs. bias that we subtract off in Fig. 3b of the main paper; however we find that the dominant contribution to the frequency-symmetric component of the ST-FMR signal is symmetric in bias. For these reasons we do not consider this final term in the main paper. The first two terms on the right in Eq. (S8) are then identical to Eq. (2) in the main text.

Equation (5) in the main text follows from Eq. (S7) after using $\omega_m = \gamma M_{eff}\sqrt{N_x N_y}$ and assuming that $\tau_\parallel(I,\theta) \propto \sin(\theta)$.

Supplementary Note 2: Details on the calibration of $I_{RF}^2$

The calibration of $I_{RF}^2$ is performed in two steps: (1) a flatness correction and (2) accounting for the bias dependence of the sample impedance. The flatness correction ensures that the microwave current within the sample $I_{RF}$ does not vary with frequency. We apply an external magnetic field $H$ with magnitude chosen so that all ST-FMR resonances have frequencies higher than the range of interest, and then measure the ST-FMR background signal as a function of frequency for a fixed DC bias ($|I| > 10\;\mu A$). Due to circuit resonances and losses, this background signal may vary as the frequency is changed. At the same time, we determine $\partial^2 V/\partial I^2$ by measuring $\partial V/\partial I$ versus $I$ with low-frequency lock-in techniques and then differentiating numerically. We can then determine the variations of $I_{RF}^2$ with frequency using the formula for the non-resonant background:

$$\langle V_{background} \rangle = \frac{1}{4} \frac{\partial^2 V}{\partial I^2} I_{RF}^2. \tag{S9}$$

We input this information to the microwave source, and employ its flatness-correction option to modulate the output signal so that the final microwave current coupled to the sample no longer varies with frequency.

(2) After step (1), $I_{RF}$ is leveled vs. frequency and its magnitude can be determined for one set of values $I_0$ and $H_0$. However, because the sample impedance varies as a function of $I$ and $H$, we must also determine how $I_{RF}$ varies as these quantities are changed. In order to do this accurately even at points where $\partial^2 V/\partial I^2$ is near zero, we calculate $I_{RF}(I,H)$ by taking into account how variations in $dI/dV$ alter the termination of the transmission line, assuming that the impedance looking out from the junction is 50 $\Omega$:

$$I_{RF}(I,H) = I_{RF}(I_0,H_0) \left[ \frac{dV}{dI}(I_0,H_0) + 50\,\Omega \right] \bigg/ \left[ \frac{dV}{dI}(I,H) + 50\,\Omega \right]. \tag{S10}$$

In practice, we generally determine $I_{RF}(I_0, H_0)$ using Eq. (S9) together with the value of the non-resonant background at one choice of $I_0$ for each value of magnetic field, and then employ (S10) to find the full $I$ dependence.

Supplementary Figure S1 shows that this procedure successfully reproduces the measured non-resonant background signal as a function of $I_0$, using as input the bias dependence of $dV/dI$ measured at low frequency. This demonstrates that there are no high-frequency phenomena which cause the background signal to deviate significantly from the simple rectification signal caused by non-linearities in the low-frequency current-voltage curve. Supplementary Figure S2 shows the typical change in $I_{RF}$ as described by Eq. (S10).

Supplementary Note 3: Regarding possible alternative mechanisms for the antisymmetric Lorentzian component of the ST-FMR signal

Kovalev et al. [S2] and Kupfershmidt et al. [S3] have noted that a component of the ST-FMR signal that is antisymmetric in frequency relative to the center frequency can arise if the precession axis of the free layer moment is tilted away from the sample plane and not along any of the principle axes of the magnetic anisotropy. In principle, this mechanism could explain an observation of an antisymmetric ST-FMR signal that varies linearly with DC current *I*, because the in-plane component of spin-transfer torque from *I* will cause the equilibrium orientation of the free-layer moment to move out-of-plane (until the torque from the demagnetization field balances the in-plane spin-transfer-torque). However, when evaluating this mechanism quantitatively, we find that it predicts an antisymmetric component 50 times smaller than we measure.

In principle, heating might affect the ST-FMR measurements through several mechanisms. Here we consider only whether a heating effect might be able to explain our observation that the ST-FMR signal contains a perpendicular component with an antisymmetric Lorentzian lineshape, whose magnitude depends approximately linearly on *I* (*i.e.*, we consider heating as an alternative mechanism to the out-of-plane torkance discussed in the main paper.) If Ohmic heating is the dominant source of heating, then the sample temperature may have an RF component proportional to $dT(t) \sim R\,(I + I_{RF}(t))^2 \sim 2RI_{RF}I\cos(\omega t + \delta_T)$ (after subtracting the constant contribution $\propto RI^2$ and assuming $I > I_{RF}$), where $\delta_T$ is a possible phase lag. If heating changes the resistance of the sample, this would give an additional contribution to the resonant part of the ST-FMR signal of the form $\langle V_{mix} \rangle \propto \frac{\partial^2 V}{\partial \theta \partial T} \langle (\delta\theta(t))(\delta T(t)) \rangle \propto \frac{\partial^2 V}{\partial \theta \partial T} I_{RF} I \,\text{Re}(m_y e^{-i\delta_T})$. However, since $\partial^2 V / \partial\theta \partial T$ in this expression is proportional to $I$ at low bias, the lowest-order contribution to the ST-FMR signal from this mechanism is proportional to $I^2$, so that it cannot explain the linear dependence of the asymmetric component on $I$ observed experimentally.

An antisymmetric-in-frequency ST-FMR signal linear in $I$ could result if the Peltier effect, rather than Ohmic heating, were the dominant heating mechanism. However, our differential conductance measurements do not show a large asymmetry with respect to bias that would be expected if this were the case. A resonant signal linear in $I$ could also result if the dominant consequence of heating were not to change the resistance, but to apply a torque to $\hat{m}$ by changing the demagnetization or dipole field. We expect that this last mechanisms might be significant if the free layer were tilted partially out of the sample plane, but we estimate that it is insignificant for our measurements in which the free-layer moment is in plane and aligned within a few degrees of the symmetry axis $\hat{z}$.

For these reasons, we believe it is unlikely that heating, rather than a direct out-of-plane spin-transfer torque, can explain the antisymmetric component of the ST-FMR signal that we observe.

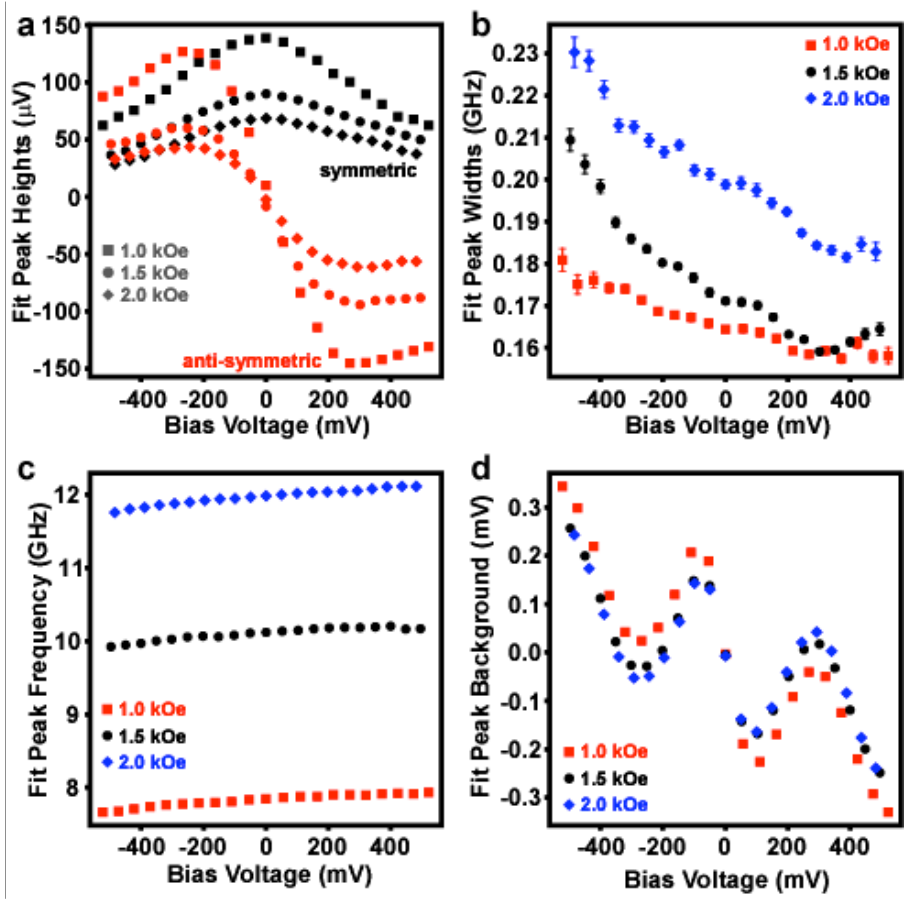

Supplementary Figure S1. Fit parameters for the ST-FMR signals at room temperature, for three values of magnetic field in the $\hat{z}$ direction and $I_{RF} \approx 12$ µA. (a) Amplitudes of the symmetric and antisymmetric Lorentzian component of each peak. (b) The linewidths $\sigma/(2\pi)$. (c) The center frequencies $\omega/(2\pi)$. (d) Non-resonant background components.

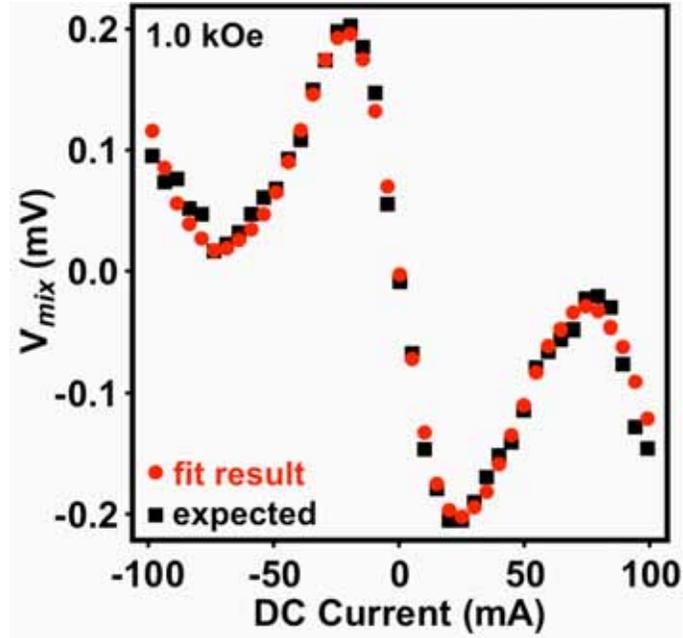

Supplementary Figure S2. Test of the calibration for $I_{RF}$ and the non-resonant background, for $H = 1.0$ kOe in the $\hat{z}$ direction. Circles: Magnitude of non-resonant background measured from fits to the ST-FMR peaks. Squares: the background expected from Equations (S9) and (S10) after determining $I_{RF} = 11.7$ μA at $I_0 = -30$ μA.

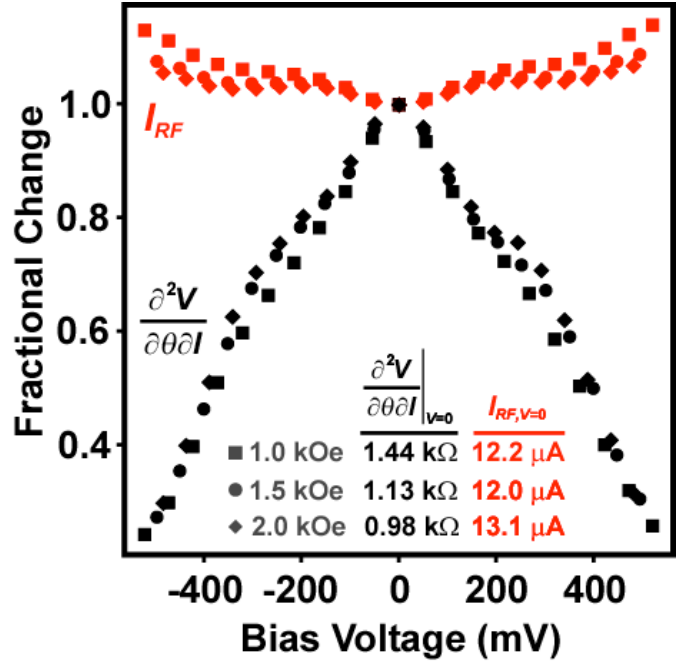

Supplementary Figure S3. Representative examples of the bias dependence of $I_{RF}$ and $\partial^2 V/\partial\theta\partial I$ for $H$ in the $\hat{z}$ direction. Values of $I_{RF}$ and $\partial^2 V/\partial\theta\partial I$ at $V=0$ are labeled. $I_{RF}^2$ is determined using the procedure described above. $\partial^2 V/\partial\theta\partial I$ is determined by measuring $\partial V/\partial I$ vs. $I$ at a sequence of magnetic fields in the $\hat{z}$ direction, by assuming that the conductance changes at zero bias are proportional to $\cos(\theta)$ and that $\theta$ depends negligibly on $I$, and then by performing a local linear fit to determine $\partial^2 V/\partial\theta\partial I$ for given values of $I$ and $H$.

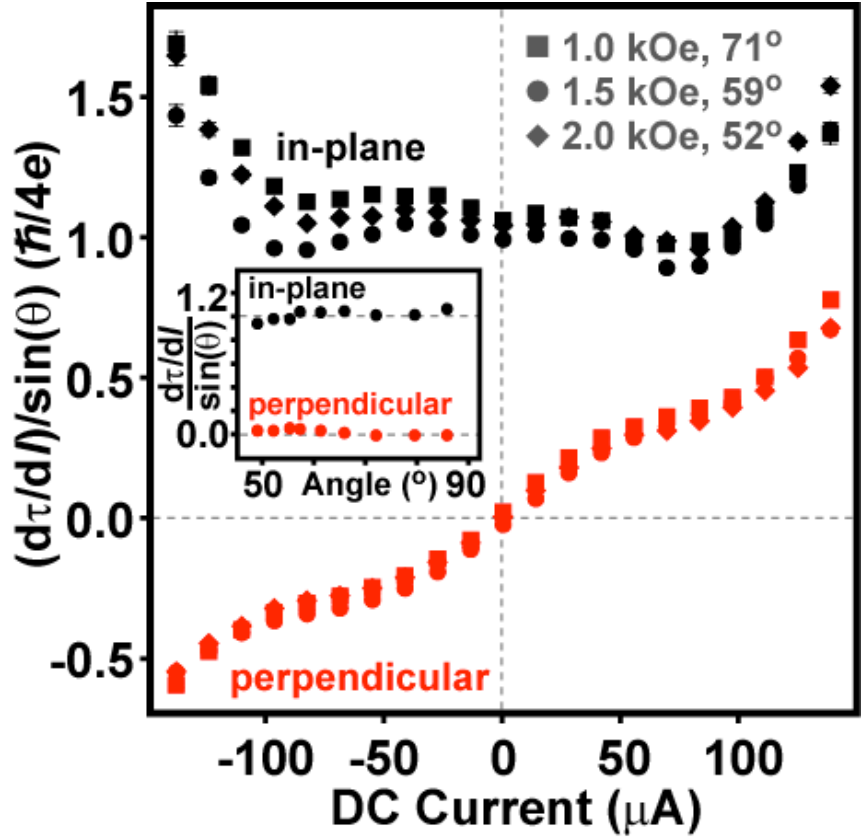

Supplementary Figure S4. Magnitudes of the in-plane and out-of plane differential torques $d\tau_{\parallel}/dI$ (black symbols) and $d\tau_{\perp}/dI$ (red symbols) vs. *I*, determined from fits to room-temperature ST-FMR spectra. The overall scale for the y-axis has an uncertainty of ~15% associated with the determination of the sample volume. (Inset) Angular dependence of the differential torques at zero bias.

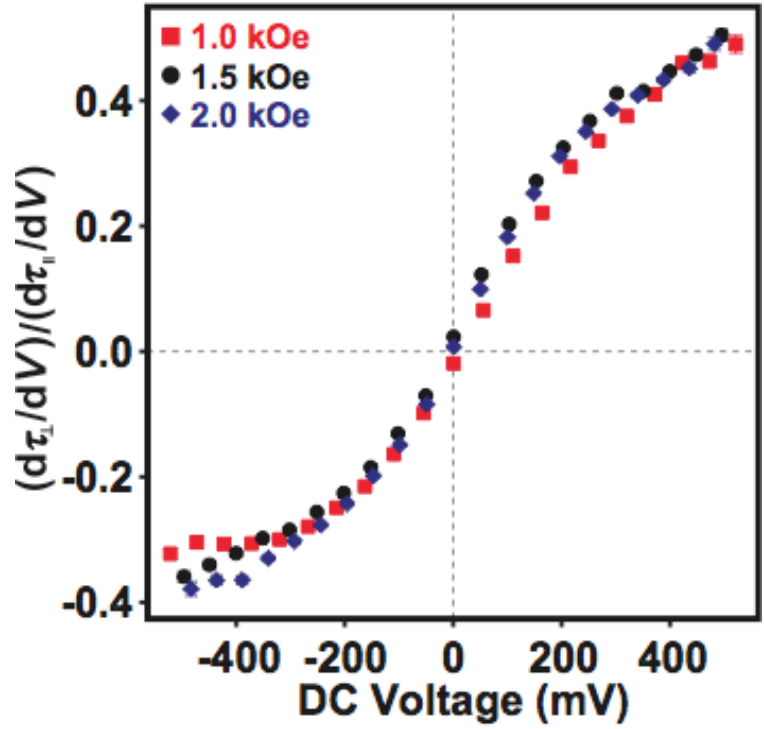

Supplementary Figure S5. Ratio of the perpendicular torkance $d\tau_\perp/dV$ to the in-plane torkance $d\tau_\parallel/dV$ as a function of bias.

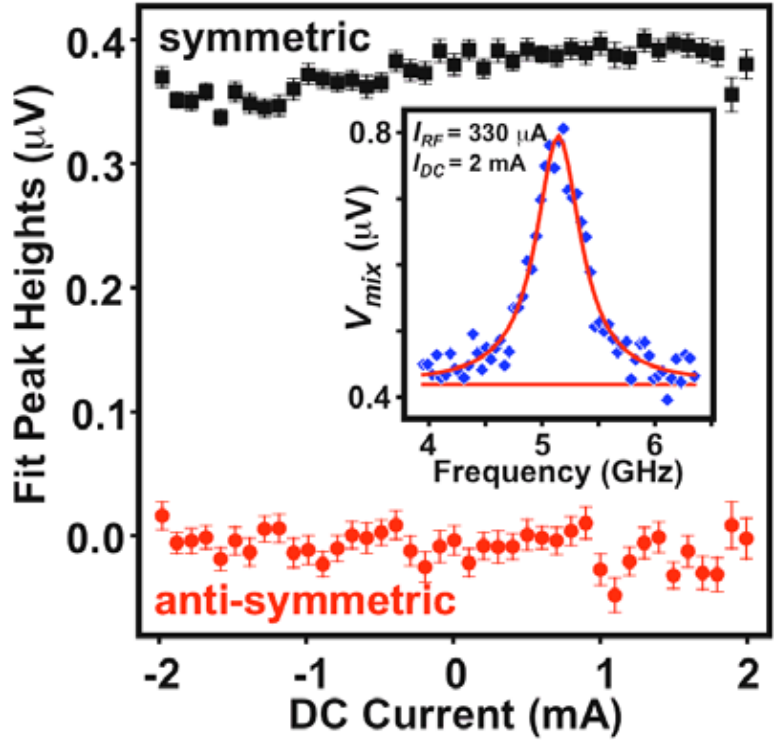

Supplementary Figure S6. ST-FMR signals for a metallic spin valve, (in nm) Py 4 / Cu 80 / IrMn 8 / Py 4 / Cu 8 / Py 4 / Cu 2 / Pt 30, with $H = 560$ Oe in the plane of the sample along $\hat{z}$ and with an exchange bias direction 135° from $\hat{z}$. We estimate $\theta = 77°$ from the GMR. The average anti-symmetric Lorentzian component is $2 \pm 3\%$ the size of the symmetric Lorentzian component over this bias range. Accounting for the out-of-plane anisotropy $4\pi M_{eff} \sim 1$ T in Eq. (2) of the main paper, we estimate that the ratio $\tau_\perp / \tau_\parallel < 1\%$.